\documentclass[usegraphicx,useAMS,usenatbib,fleqn]{mn2e}
\newcommand{\be}{\begin{equation}}
\newcommand{\ee}{\end{equation}}
\newcommand{\bd}{\begin{displaymath}}
\newcommand{\ed}{\end{displaymath}}

\usepackage{times}
\usepackage{amsmath}
\usepackage{amsfonts}

\title[Spots on PMS stars]
 {The effect of starspots on the radii of low-mass pre-main sequence stars}

\author[R. J. Jackson and R. D. Jeffries]
  {R. J.~Jackson and R. D.~Jeffries\\
   Astrophysics Group, Research Institute for the Environment, Physical
  Sciences and Applied Mathematics, Keele University, \\ Keele, 
      Staffordshire ST5 5BG
}
\date{2nd April 2014, MNRAS in press}

\pagerange{\pageref{firstpage}--\pageref{lastpage}} \pubyear{2013}

\def\LaTeX{L\kern-.36em\raise.3ex\hbox{a}\kern-.15em
    T\kern-.1667em\lower.7ex\hbox{E}\kern-.125emX}

\begin{document}
\label{firstpage}
\maketitle

\begin{abstract}
A polytropic model is used to investigate the effects of dark
photospheric spots on the evolution and radii of magnetically active,
low-mass ($M<0.5\,M_{\odot}$), pre-main sequence (PMS) stars. Spots slow
the contraction along Hayashi tracks and inflate the radii of PMS stars
by a factor of $(1-\beta)^{-N}$ compared to unspotted stars of the same
luminosity, where $\beta$ is the equivalent covering fraction of dark
starspots and $N \simeq 0.45 \pm 0.05$. This is a much stronger
inflation than predicted by the models of Spruit \& Weiss (1986) for main sequence
stars with the same $\beta$, where $N \sim 0.2$--0.3. These models have
been compared to radii determined for very magnetically active K- and
M-dwarfs in the young Pleiades and NGC~2516 clusters, and the radii of
tidally-locked, low-mass eclipsing binary components. The binary
components and ZAMS K-dwarfs have radii inflated by $\sim 10$ per cent
compared to an empirical radius-luminosity relation that is defined by
magnetically inactive field dwarfs with interferometrically measured
radii; low-mass M-type PMS stars, that are still on their Hayashi
tracks, are inflated by up to $\sim 40$ per cent. If this were
attributable to starspots alone, we estimate that an effective spot
coverage of $0.35 < \beta < 0.51$ is required. Alternatively, global
inhibition of convective flux transport by dynamo-generated fields may
play a role. However, we find greater consistency with the starspot
models when comparing the loci of active young stars and inactive field
stars in colour-magnitude diagrams, particularly for the highly
inflated PMS stars, where the large, uniform temperature reduction
required in globally inhibited convection models would cause the stars
to be much redder than observed.

\end{abstract}


\section{Introduction}
There is increasing evidence that the radii of fast rotating,
magnetically active K- and M-dwarf stars are inflated relative to the
predictions of evolutionary models. Measurements of
eclipsing binaries suggest that radii can be 10--15 per cent larger
than expected, at a given mass, for binary components with
$M<0.7\,M_{\odot}$ (Lopez-Morales 2007; Morales et al. 2009; Torres
2013). The components of these close binary pairs are expected to be
tidally locked and fast-rotating, hosting strong, dynamo-generated
magnetic fields. It has been suggested that the relative increase in
radius is associated with this magnetic activity (Lopez-Morales 2007),
although it is difficult to find slowly rotating, magnetically inactive eclipsing
binaries with which to test this hypothesis.

The same comparison between observation and theory cannot easily be
made for single stars since a direct measurement of their masses
is not possible. However, it is possible to compare the radii and
luminosities of nearby field
stars with the predictions of evolutionary
models.  Boyajian et al. (2012b, hereafter BM12) reported the
interferometric angular diameters of K- and M-dwarfs with precise
Hipparcos parallaxes and used these to determine an empirical
radius-luminosity relation for main sequence (MS) stars over the
temperature range 3200--5500\,K. The radii of these relatively inactive
field stars show satisfactory agreement with the predicted radii of
evolutionary models for K- and early M-dwarfs (e.g. the BCAH98
model of Baraffe et al. 1998) but for later M-dwarfs (spectral types M2
to M4), models generally underestimate radii (by $\sim 8$ per
cent in the case of the BCAH98 model; $\sim 5$ per cent for the
Dartmouth models of Dotter et al. 2008).  No such comparisons have been
made for magnetically active stars since there are no very active field
M-dwarfs close enough to allow a precise interferometric radius measurement.

Jackson, Jeffries \& Maxted (2009) estimated the mean
radii for groups of rapidly rotating, highly magnetically active, late
K- and M-dwarfs in the young open cluster NGC 2516 (age $\simeq
140$\,Myr -- Meynet,
Mermilliod and Maeder 1993) using the product of their rotation periods
and projected equatorial velocities. They found that their mean radii at
a given luminosity are larger than model predictions and also larger
than the interferometric radii of magnetically inactive dwarfs. The
discrepancy rises from $\sim 10$ per cent for late K-dwarfs to as much
as $\sim 50$ per cent for M4 dwarfs that are nearly or fully convective. This
result supports the suggestion that the radius inflation (at a given
luminosity) is due to rotationally induced magnetic activity. 
Further measurements are required to correlate radius inflation
with rotation, magnetic field strength and other magnetic
activity indicators, but there is sufficient evidence of a causal link to have
stimulated theoretical studies. Two mechanisms related to
dynamo-generated magnetic fields have been suggested that
could inflate the radii of K- and M-dwarfs: 
inhibition or stabilisation of convection
(e.g. Mullan \& MacDonald 2001; Chabrier et al. 2007; Feiden \& Chaboyer
2012, 2013a,b; MacDonald \& Mullan 2013); or the effect of cool,
magnetic starspots (Spruit 1982; Spruit \& Weiss 1986; Chabrier et al. 2007).

Mullan and MacDonald (2001) modelled the effect of interior magnetic
fields by modifying the Schwarzschild criterion and suppressing
convection. Chabrier et al. (2007) inhibited the convective
efficiency by artificially reducing the convective mixing-length
parameter. Feiden \& Chaboyer (2012, 2013b) described a modification
to the Dartmouth evolutionary code that takes account of the effect of
magnetic field on the equation of state and on mixing-length theory. In
their papers they give examples that produce the modest (10 per cent)
observed radius inflation of the components of several eclipsing
binaries of mass $0.4<M/M_{\odot}<1.0$, although the predicted surface
magnetic fields are a few times higher than observational
estimates. This explanation of radius inflation becomes less plausible
for lower mass stars where convective heat transfer is more and
more efficient. Either much larger changes in effective mixing length
are required to produce even a 10 per cent radius inflation in
nearly or fully convective stars (e.g. the binary CM Dra), or interior
magnetic field strengths would need to approach 50\,MG to sufficiently
stabilise the star against convection, which are considered too high to
be physically plausible (Feiden \& Chaboyer 2013a).

An alternative mechanism is the effect of starspots in reducing heat
flux out of the photosphere. Spruit \& Weiss (1986, hereafter SW86),
modelled the effect of starspots on zero age main sequence (ZAMS) stars
in the mass range 0.2--1.9\,$M_{\odot}$. They argued that the radius,
luminosity and temperature of the unspotted surface would adapt on the
Kelvin-Helmholtz timescale of the convective envelope to achieve
thermal equilibrium. For low-mass stars with deep convective envelopes they
used a polytropic model to show that the effect of starspots is to
reduce the luminosity of the star at near-constant radius and
temperature of the unspotted photosphere, hence lowering the {\it
  effective} temperature. For higher mass stars, where the mass of the
convective envelope is a small fraction of the stellar mass, SW86 used
a numerical model to show that the effect of starspots would be to
increase the temperature of the unspotted photosphere while the radius
and luminosity are nearly unchanged. 

Which of these mechanisms is dominant is undecided and likely to depend
on the fraction of the photosphere covered by dark
starspots. The presence of starspots on magnetically active stars is a
well documented phenomenon. Evidence for dark starspots and their
associated magnetic fields came initially from observations of
rotational modulation of broadband fluxes (e.g. Hall 1972, Eaton \&
Hall 1979) but there is now a large literature that detects and
investigates starspots using direct and indirect observational
techniques such as Doppler imaging (Collier-Cameron \& Unruh 1994;
Strassmeier 2002), spectroscopy of Zeeman-broadened lines (e.g. Marcy
1982, Johns-Krull \& Valenti 1996), and tomography using circularly
polarized light (Zeeman Doppler Imaging - Semel 1989, Donati et
al. 1997). However, the relevant question here is whether spots cover
enough area of the surface to significantly reduce the stellar
luminosity? Spot coverage and spot temperatures have been determined
for very active G- and K-stars from their optical TiO absorption bands,
indicating filling factors of 20--50 per cent, with temperature ratios
between spotted and unspotted photosphere of 0.65--0.76 (O'Neal, Neff
\& Saar 1998; O'Neal et al. 2004; O'Neal 2006). If starspots are
present at similar filling factors on active K- and M-dwarfs then this
could significantly reduce their luminosity at a given radius.  


MacDonald and Mullan (2013) compared measurements of (projected) radii
measured on young active M-dwarfs in NGC\,2516 (Jackson et al. 2009)
with their numerical models that include a prescription for magnetic
suppression of convection and cool starspots.  They found that, for
models in which only the effects of spots are included, the mean radii
of the coolest stars in the sample would require a spot coverage of
more than 79 per cent of the photosphere. As they considered this
implausible they rejected spots as the main cause of radius
inflation. Their estimate of the increase in radii was made by
comparison to a model for the radii of inactive stars and, as we show
in this paper, a significantly lower
estimate of filling factor is obtained if the radii of active stars are
instead compared to an empirical radius-luminosity relation for
inactive main sequence stars (e.g. from BM12).

SW86 estimated the effect of starspots only for ZAMS stars, where the
luminosity depends on the temperature of their nuclear burning
cores. The cool M-dwarfs in NGC\,2516, which show the largest apparent
increase in radius relative to inactive stars, are still in their
pre-main sequence (PMS) phase. They are likely to be fully convective
with luminosity produced chiefly by the release of gravitational energy,
and might respond differently to starspots. In this paper we: (a)
extend the work of SW86 to include the effect of starspots on low-mass
PMS stars; (b) use the recently reported radius-luminosity relation of
BM12 for inactive MS stars as a baseline against which to assess the 
increase in radius of active K- and M- dwarfs; and (c) compare these
models with the published radii of eclipsing binaries, and with the radii
of highly active K- and M-dwarfs, including a new analysis of the projected
radii of stars in the young Pleiades cluster.  We find that PMS stars
are more inflated for a given level of spot coverage than ZAMS stars of
a similar luminosity and that the spot coverage required to produce
the measured radii is large but perhaps not
excessive. Finally it is shown that, in contrast to models that
suppress convective flux and uniformly lower the photospheric
temperature, the spot model is able to simultaneously match the
loci of very active stars in colour-magnitude diagrams.

\begin{figure*}
	\centering
	\begin{minipage}[t]{0.95\textwidth}
	\centering
	\includegraphics[width = 150mm]{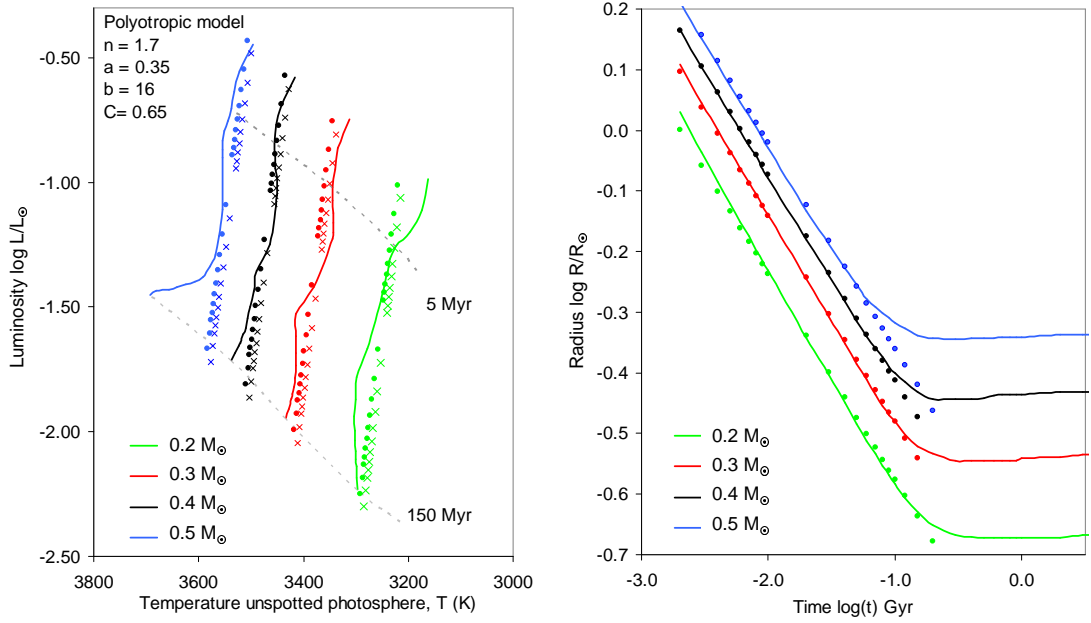}
	\end{minipage}
	\caption{ Evolutionary tracks for, 0.2, 0.3, 0.4 and 0.5 solar
          mass stars over the age range 5 to 150Myr. The solid lines in
          the left hand plot show evolutionary tracks from the BCAH98
          model of stellar evolution ($[$M/H$]$=0,  Y=0.275,  $L_{mix}$
          =$H_P$). The filled circles show the evolutionary tracks
          predicted using a polytropic model of an unspotted star on
          the Hayashi track with constants adjusted to match
          approximately the BCAH98 tracks. The crosses show the effect
          of 30 percent coverage of dark starspots for the same
          polytropic model. The right hand plot shows the variation of radius with time. The lines show the BCAH98 model and the filled circles show results of the polyropic model of an unspotted star.}
\label{fig1}
\end{figure*}

\section{The radius inflation produced by starspots on low-mass PMS stars}

\subsection{A polytropic model of PMS stars}

The effect of starspots on the radius, luminosity, and temperature of PMS stars is determined using a polytropic model of the star in the Hayashi zone (Prialnik 2000). Taking the simplest of the formulations given by SW86, where the effect of starspots is solely to inhibit radiation from the photosphere, reducing the luminosity of the star to
\begin{equation}
L=4 \pi \sigma\, (1-\beta)\,R^2\, T^4 \, ,
\label{eqn1}  	
\end{equation}
where $R$~is the radius , $T$~the temperature of the unspotted photosphere, and $\sigma$~is Stefan's constant. $\beta$ is the equivalent filling factor of completely dark starspots that would produce the same luminosity reduction as the actual coverage of starspots at the real (non-zero)  starspot temperature. 
 
Prialnik (2000, Eqns. 8.11 and 8.13) gives the following relation
between the luminosity of an unspotted star, $L_{u}$, on the Hayashi
track, as a function of $T$ and mass, $M$.
\begin{equation}
\log L_{u} =A\log T +B\log M + \rm{constant},
\label{eqn2} 
\end{equation}
where
\begin{eqnarray}
A & = & \frac{(7-n)(a+1)-4-a+b}{(3-n)(a+1)/2-1}\ \ {\rm and} \nonumber \\
B & = & -\frac{(n-1)(a+1)+1}{(3-n)(a+1)/2-1}. \nonumber
\end{eqnarray}
Here, $n$ is the polytropic index and the exponents $a$ and $b$
describe the rate of change of the Rosseland mean opacity as a function
of surface density, $\rho$, and $T$, as $\kappa=\kappa_0 \rho^a
T^b$. Rearranging this expression and reducing the luminosity to
represent the effects of starspots (i.e. $L = (1-\beta)\,L_u$) gives the
following expression for the luminosity of a spotted polytropic star;
\begin{equation}
\frac{L}{L_{\odot}}=C^{-A}(1-\beta)\left(\frac{M}{M_{\odot}}\right)^B\left(\frac{T}{T_{\odot}}\right)^A,
\label{eqn3} 
\end{equation}
where $C$~is a constant that depends on $n, a, b$~and $\kappa_0$.

The luminosity of a star on the Hayashi track results from the release
of gravitational energy as its radius contracts. The gravitational
potential energy of a polytropic star is given by $\Omega =-\tfrac{3}{5-n}GM^2 /R$~ (Hayashi, Hoshi and Sugimoto 1962). Applying the virial theorem gives the luminosity, $L=-\tfrac{1}{2}\tfrac{\partial \Omega}{\partial t}$, as a function of time, $t$;
\begin{equation}
	\frac{L}{L_{\odot}}= -\frac{3}{(10-2n)}\,\left(\frac{GM_{\odot}^2}{L_{\odot}R_{\odot}}\right)
        \left(
        \frac{M}{M_{\odot}}\right) ^2 \left( \frac{R}{R_{\odot}}\right)
        ^{-2}\frac{\partial(R/R_{\odot})}{\partial t}\, .
\label{eqn4} 
\end{equation}
 Eliminating, $L$~and $T$~from Eqns.~2 and 3 gives the rate of change of radius with time as;
\begin{equation}
\frac{d(R/R_{\odot})}{dt} = -Z(1-\beta ) \left( \frac{4-A}{4-3A} \right) \left( \frac{R}{R_{\odot}}\right)
 ^\frac{8-4A}{4-A} \left( \frac{M}{M_{\odot}}\right)
^{\frac{2A+4B-8}{4-A}}\, ,
\label{eqn5} 
\end {equation}
\noindent{where the constant  
\begin{displaymath}
Z =
  \frac{(10-2n)(4-3A)}{3(4-A)}\left(\frac{L_{\odot}R_{\odot}}{GM_{\odot}^2}\right)
  C^{\frac{-4A}{4-A}}\, .
\end{displaymath}
Integrating this expression and using the boundary condition of infinite radius at zero time gives the radius of an unspotted polytropic star as a function of time;
\begin{equation}
\frac{R}{R_{\odot}}=\left( Zt \right) ^{\frac{A-4}{4-3A}} \left(
\frac{M}{M_{\odot}} \right) ^{\frac{8-2A-4B}{4-3A}}\, .
\label{eqn6} 
\end{equation}
If spots with an effective filling factor $\beta$ develop at a time $t_s$~then the radius at time $t > t_s$~is given by
\begin{equation}
\frac{R}{R_{\odot}}=[Zt( 1-\beta) + Zt_s\beta] ^{\frac{A-4}{4-3A}} 
\left( \frac{M}{M_{\odot}} \right)^{\frac{8-2A-4B}{4-3A}}\, ,
\label{eqn7} 
\end{equation}
and the resultant luminosity is given by
\begin{equation}
\frac{L}{L_{\odot}}=(1-\beta)C^{\frac{-4A}{4-A}}[Zt(1-\beta) + Zt_s\beta] ^{\frac{2A}{4-3A}} 
\left( \frac{M}{M_{\odot}} \right) ^{\frac{4B-4A}{4-3A}}\, .
\label{eqn8} 
\end{equation}


Figure 1 compares the evolutionary tracks and radius as a function of
time predicted by the BCAH98 model (for $[$M/H$]$=0,  Y=0.275 and
$L_{\rm mix}$ =$H_P$) with the results of the polytropic model for
parameter values $A =-132$, $B =14$ and $n=1.73$ (corresponding to
$a=0.35$ and  $b=16$ in Eqn.~2). 
The constants of the polytropic model are set to match the average properties of the BCAH98 model for masses of 0.2, 0.3 and 0.4 $M_{\odot}$ over the age range 20 to 100\,Myr. Exponents $A$ and $B$ are scaled to match the slope of radius versus time (Eqn. 7) and the rate of change of $T$ with $M$ (Eqn.~3). Constant $C$ is set to match the average temperature as a function of $L$ in the Hertzsprung-Russell (HR) diagram and $n$, is chosen to match the average radii as a function of time to the BCAH98 values (Eqn.~7). Constants $a$ and $b$ which scale the mean Rosseland opacity are then found from Eqn.~2.

There is good agreement (see Fig.~1) between this polytropic model and
the BCAH98 model for $M \leq 0.4\,M_{\odot}$, indicating that these
lower mass stars are effectively PMS stars following Hayashi tracks at
ages up to $\sim 150$\,Myr. The point where stars diverge from the
BCAH98 evolutionary tracks coincides with where they leave the Hayashi
track and so the model also works at higher masses, but for a more
limited age range; a 0.5\,$M_{\odot}$~star leaves the Hayashi track at
an age of $\sim$70\,Myr. The parameters $a$~and $b$~defining the
Rosseland opacity change significantly with temperature (Alexander and
Ferguson 1994), but it is shown in section 2.2 that the predictions
made using the polytropic model are insensitive to the adoption of
constant values for these parameters provided stars follow
near-vertical tracks in the HR diagram (i.e $|A| \gg 1$), which they do
as long as they remain on Hayashi tracks.

\subsection{The effects of starspots on the radius and luminosity of
  pre-main sequence stars}

The polytropic model can be used to predict the effects of starspots on
the radius, luminosity and temperature of PMS stars. Figure 2 shows the
effects of an intentionally unrealistic case where dark starspots
\textit{suddenly} develop at a time $t_s=10$\,Myr to cover 30 per cent
($\beta$=0.3) of the surface of a 0.3\,$M_{\odot}$ star. The immediate
effect is to reduce $L$ by 30 per cent with no change in R or T. The
effect of spots is then to slow the subsequent rate of descent along
the Hayashi track by a factor $(1-\beta)$. As the star contracts, $R$ 
follows Eqn.~7, and $L$ follows Eqn.~8 such that
after $\sim 1$~dex in $\log t$, $R$ approaches the limiting radius of a
star with $\beta$=0.3 and $t \gg t_s$ (Eqn.~6), and the luminosity
converges towards its limiting value for $t \gg t_s$.  The short term
effect of an immediate reduction in luminosity is the same as that
described by SW86 for spots on low-mass MS stars with deep convective
envelopes; a reduction in $L$
with no change in $R$~and $T$. However, the longer term behaviour is
different; spotted PMS stars show an increase in $R$ relative to an
unspotted star and a reduction in $L$ of equal proportion (see below) with only a
minimal change in $T$.

The polytropic model provides simple relationships for
the increase in $R$ and reduction in $L$ of spotted stars on the Hayashi
tracks when the age of the star is much greater than the age when
starspots developed. The limiting values of $R$ and
$L$ for $t \gg t_s$ in Eqns.~7 and~8 can be used to compare the
luminosity and radius of a spotted star (suffix s), with filling factor
$\beta$, to those of an unspotted star (suffix u) of the same mass and age;
\begin{equation}
\left( L_s/L_u\right)_{M,\,t} = (1-\beta)^D  \ {\rm and} \ \left(
R_s/R_u\right)_{M,\,t} = (1-\beta)^{-D} \, ,
\end{equation}
where $D =(A-4)/(3A-4)$. 

For large $A$~the exponent $D$~tends to $1/3$. Hence the effect of starspots is to reduce the luminosity of the star while increasing its radius as the $\sim  1/3$~power of the area of unspotted photosphere.
Since the mass of single stars cannot be measured directly it is more
useful to estimate the effect of starspots on R and T at a fixed
$L$ and $t$. In this case the ratios of $R$ and $T$ for spotted and
unspotted stars are;
\begin{equation} 
\left( R_s/R_u\right)_{L,\,t} = (1-\beta)^{-N} \ \rm{and} \
\left( T_s/T_u\right)_{L,\,t} = (1-\beta)^{-(1/4-N/2) } \, ,
\end{equation}
where $N=(A-4)/(2A-2B)$. Note that the temperatures here refer to the
{\it unspotted} part of the photosphere and, for the spotted star,
this is {\it not} the effective temperature. 

\begin{figure}
	\centering
	\includegraphics[width = 75mm]{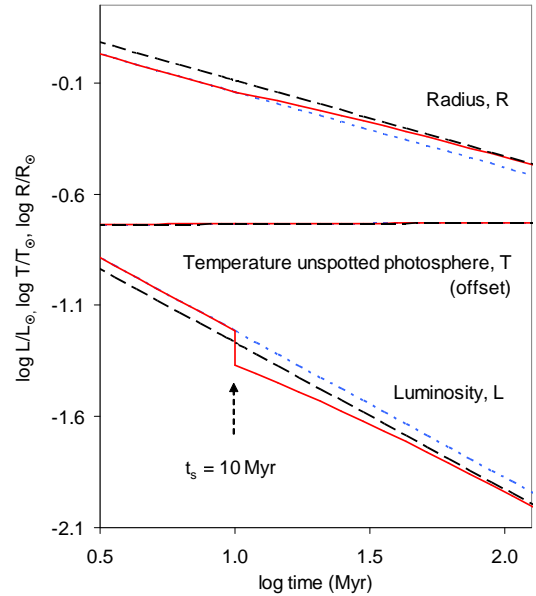}
	\caption{The predicted effects of the sudden inception of 30
          per cent coverage by dark starspots on the luminosity, radius
          and temperature of a PMS star. The lower set of curves show
          the luminosity as a function of time. The dotted curve shows
          the track of an unspotted star, the solid line shows the
          effect of spots developing at an age $t_s=10$\,Myr. The
          dashed line shows the limiting effect of spot coverage
          developing much earlier in time ($t \gg t_s$). The central
          set of curves shows the equivalent changes in the temperature
          of the unspotted photosphere (which are offset by $-0.5$ for clarity). The upper curve shows the corresponding changes in radius. }

\label{fig2}
\end{figure}

The exponents in these expressions are constant for a given polytropic
model, depending primarily on $A$, the slope of the evolutionary tracks
in the HR diagram. Thus, if the starspots were formed at a sufficiently earlier time ($t \gg t_s$), then the relative change in radius is independent of luminosity whilst the star remains on the Hayashi track.

Figure 3 shows the increase in $R$ and $T$ as a function of $\beta$~ for $N
= 0.47$, calculated using the values of $A$ and $B$ that provided
the best fit to the BCAH98 model in Section 2.1, which themselves
depend on $a$ and $b$ and which are probably dependent on $T$. To assess the
sensitivity of $N$ to changes in the parameters defining the polytropic
model, values of $a$ and $b$ were instead determined from the tables of low
temperature opacities (at X=0.70 and Z=0.02) of Alexander and Ferguson
(1994) for $3160 \leq T \leq 4000$\,K  and 
$0.001 \leq \rho \leq 0.004$\,kg\,m$^{-3}$, representative of the photospheric
properties of stars of mass $0.2< M/M_{\odot}<0.5$ and ages 10--150\,Myr
(Siess, Dufour \& Forestini 2000). This yields a range of values $0.23< a < 0.53$ and
$2.4 < b < 7.2$, resulting in an exponent $N$ increasing from 0.38 for
the youngest, lowest mass stars to 0.45 for the older stars (for a
polytropic index $1.5 < n < 2.0$). Fig.~3 shows the relative increase
in $R$ and $T$ as a function of $\beta$ evaluated using these upper and
lower bounds for $N$. These are quite similar to the best fit,
indicating that for a practical range of input parameters ($n$, $a$ and
$b$), the prediction of the polytropic model is that starspots will
principally increase the radius of a PMS star (at a given luminosity
and age) with a much lower relative increase in temperature, (the index
$(1/4-N/2) < 0.06$). In the limit, as $\partial L/\partial T$ tends to infinity , $N$ tends to 0.5 (see Eqn.~10), in which case $R$ increases with spot coverage as $R_s/R_u = (1-\beta)^{-1/2}$  and $T$ is unchanged.

\begin{figure}
	\centering
	\includegraphics[width = 80mm]{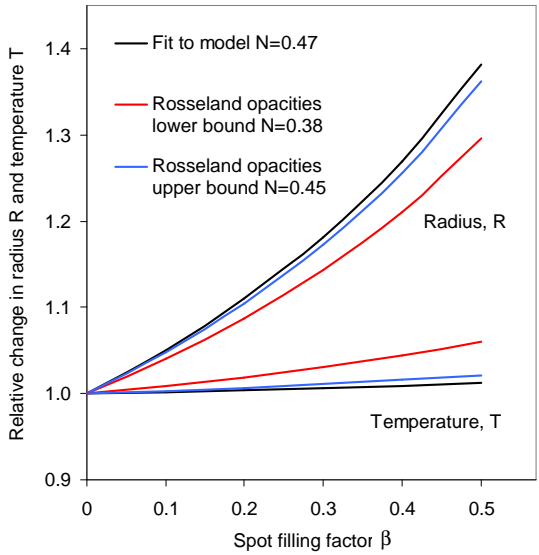}
	\caption{ Variation in radius, $R$ and temperature of the
          unspotted photosphere, $T$ for PMS stars as a function of
          spot filling factor, compared to an unspotted star of
          equivalent age and luminosity. Results are shown for
          different values of polytropic index, $n$ and parameters $a$
          and $b$ defining the Rosseland mean opacity ($\kappa =\kappa
          _0 \rho ^a T^b$). The first set shows parameters $a$, $b$ and
          $n$ that give the best fit to the BCAH98 evolutionary model (see Fig.~1). The remaining curves show the upper and lower bounds calculated for a practical range of parameter values (see section~2.2).}
          \label{fig3}
\end{figure}

\subsection{The effects of starspots on the radius and luminosity of
  main sequence stars}

Figure 4 shows the variation of the exponent $N$ with luminosity for MS
stars using the results of SW86. Figure 8 of SW86 shows the rate of change of $L$, $R$~and
$T$~with $\beta$ for stars of mass  $0.2<M/M_{\odot}<1.9$. For the
lowest mass stars the effect of spots is to reduce $L$ with only small
increases in $R$~and $T$. As the stellar mass increases, the reduction
in $L$~falls to zero and instead $R$~and $T$~increase with
$\beta$. Above $\sim 0.8M_{\odot}$~the depth of the convective envelope
becomes comparable with the predicted depth of inhibited convection
below the starspot producing a more complex pattern of changes in $L$,
$R$~and $T$. SW86 showed that for a $1M_{\odot}$ star, the changes in $R$
and $T$ scaled almost linearly with filling factor for $\beta \le
0.12$. In our paper the radii of spotted MS stars at fixed luminosity
are scaled  as $(1-\beta)^{-N}$ and are extrapolated to higher filling
factors than were quoted by SW86. Figure 4 indicates that MS stars show
markedly different changes in $R$ and $T$ with $\beta$ compared to PMS
stars on the Hayashi track. The exponent $N$ determining the increase
in radius in PMS stars is significantly larger and is independent of
luminosity. A principle result of this paper is that a given
spot coverage increases the radius (at fixed luminosity) by far more
for PMS stars than for MS stars.

\begin{figure}
	\centering
	\includegraphics[width = 75mm]{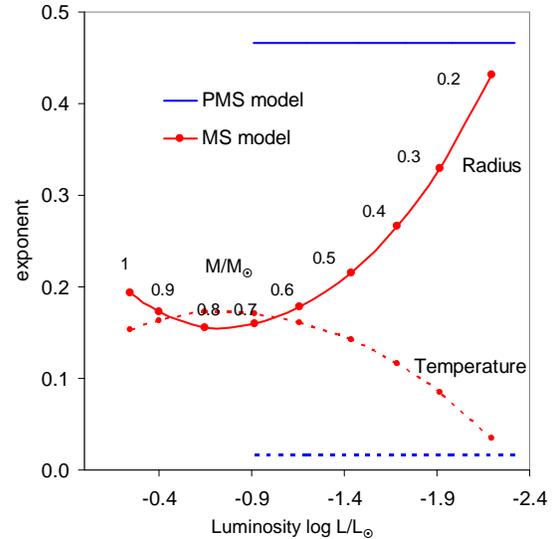}
	\caption{ Variation of the exponent $N$, which determines the
          increase in radius, $R$~as a function of filling factor of
          dark starspots, $\beta$ for stars of fixed age and luminosity
          ($R_{s}/R_{u} \propto (1-\beta)^N$). The curve for MS stars is derived
          from the results of SW86. The horizontal line for PMS stars
          shows the result for a polytropic model of stars on the
          Hayashi track (see Section 2.2). Filled circles on the MS
          curve indicate model-dependent masses of an unspotted star
          (from BCAH98). The corresponding temperature exponent,
          $1/2-N/4$ gives the increase in temperature of the unspotted
          part of the
          photosphere ($T_{s}/T_{u}\propto (1-\beta)^{1/2-N/4}$, shown as dashed lines).}
\label{fig4}
\end{figure}

\section{Empirical radius-luminosity relations for K- and early M-dwarfs}

In this section we compare the measured radii of stars from three sources:
\begin{itemize}
	\item Interferometric radii measured for relatively inactive field stars. These provide a model-independent baseline for comparison with fast-rotating, active stars.
	\item The average projected radii of fast rotating (projected
          equatorial velocities, $v\sin i > 8$\,km/s), and therefore magnetically active, low-mass stars in two young clusters, NGC\,2516 and the Pleiades aged 100--150\,Myr (Meynet et al. 1993).    
	\item The radii of short period, magnetically active, eclipsing
          binary components with known parallaxes and hence calculable luminosities.
\end{itemize}
Radii are compared as a function of luminosity since there is no model-independent method of measuring the mass of individual stars.

\begin{figure}
	\centering
	\includegraphics[width = 80mm]{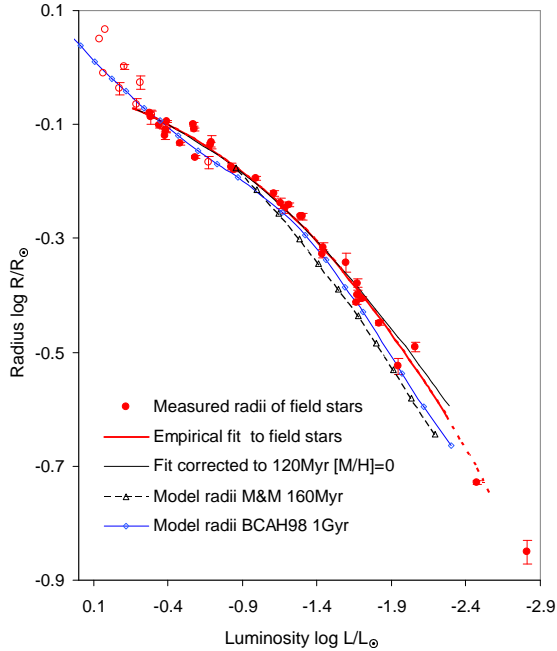}
	\caption{ The radius of unspotted K- and M- dwarfs as a
          function of luminosity. Filled circles show measured radii
          from Boyajian et al. 2012b used to define the empirical
          radius-luminosity relation (thick red line). Open circles
          show measured data for G stars from Boyajian et
          al. 2012a. The upper curve (thin black line) shows the
          empirical radius-luminosity relation "corrected" to 120\,Myr
          and [M/H]=0 as described in section 3.1. The lower curves
          show a 1\,Gyr model isochrone from BCAH98 ([M/H]=0, Y=0.282,
          $L_{mix}$=1.9 $H_P$) and the 160\,Myr isochrone used by
          MacDonald \& Mullan (2013) as their ``inactive'' baseline (see
          section 3.1).}
         \label{fig5}
\end{figure}

\begin{figure}
	\centering
	\includegraphics[width = 80mm]{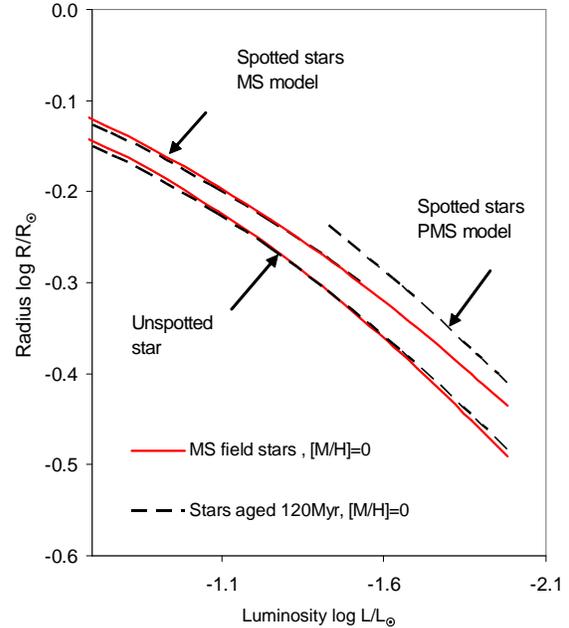}
	\caption{ The effect of dark starspots on the radius-luminosity
          relation of low mass stars. The solid curves show (a) the
          empirical radius-luminosity relation we adopt for
          magnetically inactive, solar metallicity, MS field stars (the
          lower curve) and (b) this curve modified for the
          effects of 30 per cent coverage by dark starspots, as
          predicted using the MS model (upper curve). The 
          dashed lines show (a) the empirical relation for magnetically
          inactive stars corrected to an age of 120\,Myr (the lower
          curve, almost indistinguishable from the inactive field star
          locus) and (b) the predicted effects of 30 per cent coverage
          of dark starspots using the MS model for higher luminosity
          (mass) stars that have left the Hayashi track at 120\,Myr
          (middle curve, almost indistinguishable from the spotted
          field star curve) and (c) the effects of 30 per cent spot
          coverage according to the PMS model, appropriate for lower
          luminosity (mass) stars on their Hayashi tracks (upper
          curve).}
\label{fig6}
\end{figure}

\subsection{The radius-luminosity relation of inactive field stars} 
BM12 used the interferometric radii of 33 MS K- and M-dwarfs to derive
an empirical relation for luminosity as a cubic polynomial of radius
over the range $-2.8 < \log (L/L_{\odot}) < -0.3$. Figure~5 shows the
resultant radius-luminosity relation  together with the measured radii
of the field stars.  Also shown as open circles are the radii of higher
mass stars from Boyajian et al.(2012a) not included in the fit. The
mean metallicity of the sample stars is [M/H]$=-0.15\pm 0.21$~and all
show relatively low levels of X-ray activity; all but one have $\log
(L_x/L_{\rm bol}) < -4.5$~with the highest being -3.5 (BM12), substantially
below the ``saturated'' level of $\log (L_x/L_{\rm bol}) = -3$ 
seen in the most magnetically
active stars (Pizzolato et al. 2003). In this paper the empirical
relation of BM12 is assumed to represent magnetically inactive MS stars 
with [M/H]$=-0.15$. 

In order to compare results with radii measured in the Pleiades and
NGC~2516 (see Section 3.2), small corrections are  added to the
empirical relation of BM12 to account for differences in the
metallicity and age of these clusters from the field star
sample. Soderblom et al. (2009) gives a metallicity of [Fe/H] =
0.03$^{+0.02}_{-.05}$~ for the Pleiades and Terndrup et al. (2002)
gives  [Fe/H] = -0.05$\pm$0.14 for NGC\,2516; i.e. both clusters have
near-solar metallicity. The radius-luminosity relation of BM12 shows no
obvious dependence on metallicity defined by a sample which has
metallicities in the range $-0.8<$[Fe/H]$<0.4$} . Any dependence is
small compared to the uncertainties in luminosity and/or radius over
this metallicity range. Nevertheless, the residuals to their polynomial
fit do show a weak correlation with metallicity (a Pearson's r
coefficient of 0.3). Fitting a regression line to these residuals gives
a small correction term such that  $\Delta \log
(L/L_{\odot}) = -0.11$([Fe/H]$-0.15$) at fixed $R$. i.e. $\Delta \log
(L/L_{\odot}) = 0.02$ for a solar metallicity cluster. To take account of
the age difference a second correction, $\Delta \log(R/R_{\odot})$, is found
from the difference between the radius as a function of luminosity predicted
at the cluster age and the radius predicted for ZAMS stars (age 1\,Gyr)
in the BCAH98 models. Results are shown in Fig.~5,  where the
uppermost curve shows the empirical radius-luminosity relation
"corrected" to solar metallicity and an age of 120\,Myr. There is a
small increase in radius at low luminosities (up to 7 per cent) but a
negligible change from the uncorrected BM12 empirical relation if $\log L/L_{\odot} > -1.5$. 

Comparing the empirical radius-luminosity curve in Fig.~5 with the
1\,Gyr, [M/H]=0.0 isochrone of BCAH98 shows an increasing discrepancy
at low luminosities ($\log L/L_{\odot}< -1.5$) where the empirical fit 
is $\sim 8$ per cent above the BCAH98 model after accounting for
the metallicity difference. This is a significant
difference. The scatter of measured radii around the empirical
relationship is about 6 per cent (see fig.12 of BM12) for low mass
stars, leading to a systematic uncertainty of 2--3 per cent in its
definition at low luminosities. The lowest curve in Fig.~5 shows the
baseline radii used by MacDonald and Mullan (2013) to represent
stars in NGC~2516 with no magnetically induced increase in radius. At
low luminosities ($-2.3<L/L_{\odot}<-1.5$) these radii are about 15 per
cent less than the empirical relation of BM12 (corrected to [M/H]=0,
120\,Myr). This difference cannot be accounted for by age uncertainties
for the cluster. The adoption of such a different zero-activity baseline
will of course lead to corresponding differences in the inferred radius
inflation and hence the levels of starspot
coverage required to explain the observed radii of active stars.

The (corrected) empirical radius-luminosity curve in Fig.~5 can now be
scaled using Eqn.~9 and the exponent values shown in Fig.~4 to predict
the effects of spot coverage on PMS and MS stars. Typical results are
shown in Fig.~6 for low mass stars of luminosity $-0.7 >\log
L/L_{\odot}> -2.0$. The solid lines shows the radius-luminosity
relation for inactive ZAMS stars and the predicted radius inflation due
to the effect of 30 per cent coverage ($\beta=0.3$) of dark starspots
{\it on a MS star}. Also shown as dashed lines are results for stars
aged 120\,Myr.  At this age the higher luminosity (and mass) stars,
with $\log L/L_{\odot}\geq -1.3$, are expected to be approaching the MS and
lower luminosity stars to be PMS. For unspotted stars and for spotted
stars on the MS the change in age produces only a small ($<$0.01\,dex)
increase in $\log R$ relative to the ZAMS. However, the effect of the
same starspot coverage on the low-mass PMS stars is predicted to be much
larger (by a factor of two) than the ZAMS stars. Hence the radius inflation as a function of luminosity and age will depend on whether stars are best represented by MS or PMS models (see section 4). 

\begin{figure}
	\centering
	\includegraphics[width = 85mm]{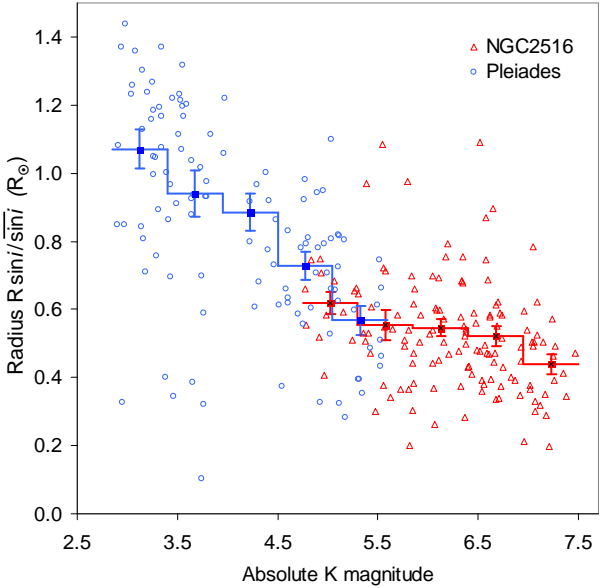}
	\caption{The mean radii of K- and M-dwarfs in NGC\,2516 and the Pleiades. Points show the projected radii of individual targets normalised to the mean $\sin i$~($R \sin i/R_{\odot}\overline{\sin i}$). Solid lines show the average radius measured in 5 equal bins for each cluster (see Table 2). }
          \label{fig7}
\end{figure}

\begin{figure}
	\centering
	\includegraphics[width = 80mm]{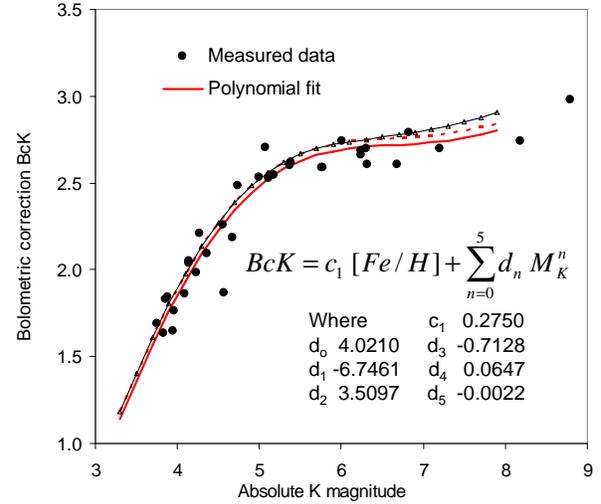}
	\caption{The bolometric correction of K- and M-dwarf stars with
          measured radii as a function of their absolute $K$
          magnitude. Filled circles show the fitted data points from
          Table 7 and 8 of BM12. The coefficients $c_1$ and $d_n$ give
          the polynomial fit to the measured data over the range
          $3.3<M_K<8.0$. The solid and dotted lines shows the
          polynomial evaluated at the mean metallicity of [M/H]$=-0.15$
          and [M/H]$=0$ respectively.  The solid line marked with small triangles shows the bolometric correction corrected to an age of 120\,Myr and [M/H]$=0$ as described in section 3.2.}
\label{fig8}
\end{figure}

\subsection{The radii of highly active stars in NGC~2516 and the Pleiades}
The radii of low mass stars in NGC~2516 and the Pleiades are
estimated from the rotational period and $v\sin i$ 
following the method described in Jeffries (2007)
and Jackson et al. (2009). The product of these quantities gives the
projected radius using the formula $R\sin i /R_{\odot}  =
0.02\,P\,v\sin i$, where $P$~is the period in days and $v\sin i$ is in
km\,s$^{-1}$. Assuming stellar spin axes are randomly oriented
(e.g. Jackson \& Jeffries 2010a), then the mean value of $R\sin i$ for
a group of similar stars can be divided by the average value of $\sin
i$ to give an estimate of the mean radius. A complication is that if
$i$ were small then it would not have been possible to measure the
periods of stars or resolve their $v\sin i$. To account for this, $\sin
i$ is averaged over $\tau < i \leq \pi /2$, where the cut-off
inclination $\tau$ is estimated by modelling the distribution of $R\sin i$
values about the mean $R\sin i$ as a function of $M_K$. Further details
are given in Jackson et al. (2009) and this correction turns out to be
small (see below).

Data for the Pleiades were taken from Table~1 of Hartman et al. (2010),
which reports rotational periods measured from results of the HATNet
survey together with $v\sin i$ values from the literature. P, $v\sin i$
and $K$ magnitudes are available for 242 stars. From these we choose a 
subset of 124 with well-resolved 
$v\sin i > 8$\,km\,s$^{-1}$. This limit is imposed
to ensure reasonably precise $R \sin i$ values and these stars are also 
expected to be highly magnetically active (see below).
Absolute $K$ magnitudes were calculated using a distance
modulus of 5.60$\pm$0.04 estimated from main sequence fitting
(Pinsonneault et al. 1998) and a reddening of E(B-V)=0.032 (An,
Terndrup \& Pinsonneault 2007). Using the relations of Reike \& Lebofsky (1985)
we estimate $A_K \simeq 0.01$ . $R\sin i$ values were averaged in 5 equal bins of
absolute $K$ magnitude between $2.85< M_K< 5.60$. The lower cut-off in
$\sin i$ was estimated as $\tau=0.26\pm 0.10$~radians which yields a
mean value of $\sin i = 0.81 \pm 0.02$ (compared to a value of $\pi/4
=0.79$ if there were no cut-off). Figure~7 shows the estimated radii for
individual targets ($R\sin i/\overline{\sin i}$) together with the mean
radii in each $M_{K}$ bin. Uncertainties in the mean are calculated
from the measured uncertainties in the mean and the uncertainty in
$\overline{\sin i}$ added in quadrature.

Data for low mass stars in NGC\,2516 were taken from Table 2 of Jackson
and Jeffries (2010b).  141 targets were selected with tabulated values
of P, $v\sin i$~and $K$ magnitude and with both $v\sin i >8$\,km\,s$^{-1}$
and greater than twice
its estimated uncertainty. Absolute $K$ magnitudes were calculated
using a distance modulus of 7.93$\pm$0.14 from main sequence fitting
and a reddening of $E(B-V)=0.12 \pm 0.02$ (Terndrup et al. 2002),
giving $A_K =0.045$. $R\sin i$ values were averaged in 5 equal bins
between $4.75< M_K< 7.5$. Two targets showing an anomalously large
$R\sin i$~(1.5\,$R_{\odot}$ and 4.0\,$R_{\odot}$) were rejected. The
lower cut-off of in inclination was estimated as $\tau$=0.40$\pm$0.08
radians, giving $\overline{\sin i} =0.83 \pm 0.02$. Estimated radii for individual targets are shown in Fig.~7 together with mean values for each bin. Where the range of luminosities sampled for the two clusters overlaps ($4.7<M_K<5.7$), the mean radii agree within their uncertainties.

To assess the likely levels of magnetic activity in these samples, 
Rossby numbers were calculated from the ratio of period to an
empirically estimated convective turnover time. This was calculated such that
coronal and chromospheric activity indicators satisfy a single scaling
law with Rossby number, independent of stellar mass (e.g. Noyes et
al. 1984). We follow the formulation of  Pizzolato et al. (2003), 
setting the turnover time to be $10^{1.1}\,(L/L_{\odot})^{-1/2}$\,days. 
The results show that all the target stars in NGC\,2516, and all Pleiades stars
with $v\sin i >$8\,km/s and $\log L/L_{\odot}<-0.3$, have $\log
N_R < -1$ and are therefore expected to have saturated levels of
magnetic activity (Pizzolato et al. 2003; Jackson et al. 2010b;
Jeffries et al. 2011). Median values of Rossby number for each of the 5
luminosity bins in each sample are given in Table 1.

\begin{figure*}
	\centering
	\begin{minipage}[t]{0.95\textwidth}
	\centering
	\includegraphics[width = 140mm]{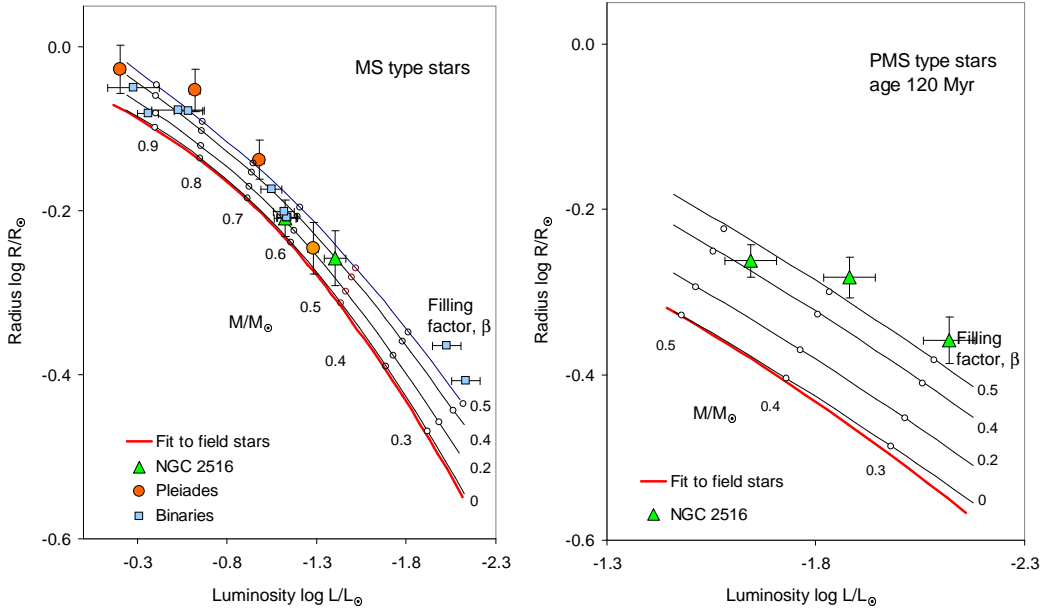}
	\end{minipage}
	\caption{Comparison of the measured radii of fast rotating low
          mass stars with the predicted radii of MS and PMS stars for
          different levels of starspot coverage. In the
          left hand panel the red solid line shows an empirical
          (magnetically inactive) ZAMS
          locus (corrected to [M/H]$=0$, see section 3.1). Above this
          are curves showing the effects of a 20, 40 and 50 per cent
          coverage by dark starspots (i.e. $\beta = 0.2, 0.4, 0.5$).  
          The right hand panel shows similar curves for PMS stars at an
          age of 120\,Myr. In both plots circles and triangles 
          show the mean radii
          measured for low-mass stars in NGC\,2516 and the Pleiades
          respectively (see section 3.2); 
          squares show the radii of eclipsing binaries
          with Hipparcos distances (see section 3.4). The curves are
          marked at intervals with model-dependent masses (in
          $M_{\odot}$) from BCAH98.}	
          
\label{fig9}
\end{figure*}

To compare the radii of cluster stars with the radii of MS field stars,
luminosities are estimated from absolute $K$ magnitudes. BM12
give luminosity as a function of various colours but not absolute
magnitude. The luminosities, $K$ magnitudes and distances reported
in Tables 7 and 8 of BM12 are used here to fit a polynomial describing
the bolometric correction, $BC_K$ against $M_K$ over the range $3.7<
M_K< 8.0$ (see Fig.~8). The Johnson $K$ magnitudes in BM12 are
transformed to the CIT system using the conversion giving in Leggett
(1992). Once again it is necessary to make a small correction
($<$0.07\,mag.) to account for the younger age and higher metallicity
of the clusters compared to the field stars. This correction is determined by comparing the values of $BC_K$ predicted using BT-Settl model atmospheres (Allard et al. 2003) at the cluster age to those at 1\,Gyr. The results are shown in Fig.~8 for an age of 120\,Myr. 

\begin{table}
	\caption{The mean radii of K- and M-dwarfs in NCG~2516 and the
          Pleiades. The average radii ($\overline{R\sin
            i}/\overline{\sin i}$) are calculated in 5 equal bins of
          $K$ magnitude for each cluster. Also shown are the number of
          targets averaged in each bin, the luminosity calculated from the average absolute $K$ magnitude (using the polynomial fit of $BC_K$ shown in Fig.~8) and the median Rossy number, $\log N_R$.}
		\begin{tabular}{lccccc}
\hline
Cluster & $M_K$ & $\log L/L_{\odot}$ & No. & $R/R_{\odot}$ & $\log N_R$ \\ \hline
Pleiades & 3.17 & 0.26+/-0.03 & 30 & 1.07+/-0.06 & -0.6 \\
Pleiades & 3.7 & -0.2+/-0.03 & 29 & 0.94+/-0.07 & -0.6 \\
Pleiades & 4.23 & -0.62+/-0.03 & 12 & 0.89+/-0.05 & -1.8 \\
Pleiades & 4.78 & -0.98+/-0.03 & 25 & 0.73+/-0.04 & -1.7 \\
Pleiades & 5.27 & -1.28+/-0.03 & 19 & 0.57+/-0.04 & -1.9 \\
NGC\,2516 & 5.09 & -1.12+/-0.06 & 13 & 0.62+/-0.03 & -1.7 \\
NGC\,2516 & 5.57 & -1.40+/-0.06 & 26 & 0.55+/-0.04 & -1.9 \\
NGC\,2516 & 6.13 & -1.64+/-0.06 & 36 & 0.55+/-0.03 & -1.9 \\
NGC\,2516 & 6.65 & -1.88+/-0.06 & 37 & 0.52+/-0.03 & -2.2 \\
NGC\,2516 & 7.12 & -2.12+/-0.06 & 25 & 0.44+/-0.03 & -2.4 \\ \hline
		 \end{tabular}
	\label{tab1}
\end{table}

\subsection{Potential systematic errors in the projected radii}
Hartman et al. (2010) discussed two reasons why there might be
systematic errors in the radii estimated from the mean value of $P\,v\sin
i$. The first is differential rotation which may cause the rate of
surface rotation to reduce towards the poles (Krause \& Raedler
1980). Using a solar-type differential rotation law, the rate of
differential rotation $\Omega = \Omega_{\rm eq} -\Delta \Omega
\sin^2 \Phi$, where $\Omega_{\rm eq}$ is the rate of rotation at the
equator, $\Delta \Omega$ is the maximum differential rotation and
$\Phi$ is the stellar latitude. Assuming
starspots are uniformly distributed then the average measured rate of
rotation will be $\Omega_m=\Omega_{\rm eq}(1-\tfrac{1}{3}\Delta \Omega)$,
so the measured period will be a factor $(1-
\tfrac{1}{3}\Delta\Omega/\Omega_{\rm eq})^{-1}$ greater than the true equatorial
period. In the present case the effect of differential rotation is
minimised by restricting the sample to relatively fast rotating stars
($v\sin i > 8$~km/s) giving a median angular velocity of 6 radians\,d$^{-1}$ 
for NGC\,2516 and 3 radians\,d$^{-1}$ for the Pleiades, compared to a typical
differential rotation of $\Delta \Omega \sim 0.07$~radians\,d$^{-1}$ estimated
for active K- and M-dwarfs by Reinhold, Reiners \& Basri (2013). Accounting
for this level of differential rotation would increase the
measured period by $<1$ per cent over the true equatorial value, but
this would be offset by a small reduction in the measured $v \sin i$ below
its true equatorial value. The net change in inferred radius ($<0.3$
per cent) is negligible compared to the observed radius inflation and
other sources of error.

A second source of error considered by Hartman et al. (2010) is
over-estimation of $v\sin i$ from the rotational broadening of spectral
lines, due to an under-estimate of the ``zero velocity'' line widths of
reference, slowly rotating stars. This possibility can be tested by
restricting the NGC~2516 and Pleiades samples to higher $v\sin i$
objects which are less sensitive to this zero-point error. Increasing
the $v\sin i$ threshold to 16\,km\,s$^{-1}$ in NGC~2516 reduces the
sample size by 20 per cent and reduces the inferred mean radii by just 1 per
cent, which is within the measurement uncertainty. The
Pleiades data, containing stars of higher mass, cannot be tested in the
same way since there are too few fast-rotating targets with $v\sin i >16$\,km\,s$^{-1}$.

\subsection{The radii of eclipsing binary stars with known distances}
A further source of measured radii as a function of luminosity are
eclipsing binary stars of known distance. Table 2 lists the parameters
of six binary stars with Hipparcos parallaxes and low-mass components
with luminosities in the range  $-0.3 > \log(L/L_{\odot}) > -2.5$. In
these cases radii are determined precisely from the eclipse light
curves. The temperature ratios are less precise, usually being
inferred from spectral type estimates. The total luminosity is
calculated from 2MASS $K$ magnitudes transformed to the CIT system
(Carpenter 2001) and using the BC$_{K}$ in Fig.~8 (assuming a
metallicity, [Fe/H]\,$=-0.15$). The $K$ magnitude of a binary is
partitioned such that the luminosity of individual components ($a$ and
$b$) is split according to their radii and temperatures (i.e. as $R_a^2
T_a^4/R_b^2T_b^4$).  The uncertainty in luminosity is calculated from
the uncertainty in parallax and the uncertainty in BC$_K$ summed in quadrature. There will be additional systematic errors if the actual metallicity of these stars differs significantly from the mean field star value used to calculate BC$_K$.

\begin{table*}
	\caption{ The radii and luminosities of binary star
          components. Luminosities are calculated from the 2MASS $K$
          magnitude (Cutri 2003) and Hipparcos parallaxes (van Leeuwen
          2007) as described in section 3.4. Masses, radii and
          temperature ratios are from Torres, Andersen and Gimenez (2010) except for AE For where data are from Rozyczka el al. (2013).}

\begin{tabular}{lllllll}
\hline
Identifier & CG Cyg & YY Gem & CU Cnc & UV Psc & V 636 Cen & AE For \\ \hline
$M_a/M_{\odot}$ & 0.941 +/- 0.014 & 0.599 +/- 0.005 & 0.435 +/- 0.001 & 0.983 +/- 0.008 & 1.052 +/- 0.008 & 0.631 +/- 0.004 \\
$M_b/M_{\odot}$ & 0.814 +/- 0.013 & 0.599 +/- 0.005 & 0.399 +/- 0.001 & 0.764 +/- 0.005 & 0.855 +/- 0.003 & 0.620 +/- 0.003 \\
$R_a/R_{\odot}$ & 0.893 +/- 0.012 & 0.619 +/- 0.006 & 0.432 +/- 0.006 & 1.110 +/- 0.023 & 1.019 +/- 0.004 & 0.67 +/- 0.03 \\
$R_b/R_{\odot}$ & 0.838 +/- 0.011 & 0.619 +/- 0.006 & 0.392 +/- 0.009 & 0.835 +/- 0.018 & 0.830 +/- 0.004 & 0.63 +/- 0.03 \\
$T_a/T_b$ & 1.114 +/- 0.041 & 1.000 +/- 0.037 & 1.011 +/- 0.068 & 1.217 +/- 0.029 & 1.180 +/- 0.029 & 1.011 \\
Parallax (mas) & 12.0 +/- 2.2 & 64.1 +/- 3.8 & 90.4 +/- 8.2 & 14.6 +/- 1.3 & 13.9 +/- 0.9 & 31.8 +/- 2.0 \\
$K_{\rm 2MASS}$ & 7.75 +/- 0.02 & 5.24 +/- 0.02 & 6.60 +/- 0.02 & 7.13 +/- 0.02 & 7.06 +/- 0.02 & 6.66 +/- 0.02 \\
$\log L_a/L_{\odot}$ & -0.28 +/- 0.15 & -1.14 +/- 0.06 & -2.03 +/- 0.08 & 0.00 +/- 0.08 & 0.11 +/- 0.06 & -1.05 +/- 0.06 \\
$\log L_b/L_{\odot}$ & -0.53 +/- 0.15 & -1.14 +/- 0.06 & -2.13 +/- 0.08 & -0.58 +/- 0.08 & -0.36 +/- 0.06 & -1.12 +/- 0.06 \\ \hline
 \end{tabular}
\label{tab2}
\end{table*}

\section{Comparing measured and predicted radii}

The measured radii in Tables 1 and 2 can be compared to the predicted
radii of spotted stars as a function of luminosity using the MS and PMS
models shown in Fig.~4. To do this it is necessary to decide which
model is appropriate for which data i.e. at what luminosity do stars in
the Pleiades and NGC\,2516 show a radius inflation characteristic of
PMS stars, of MS stars, or somewhere in between? 
This requires estimates of the cluster ages and reference to an
evolutionary model. The Pleiades has an age of 100$\pm$20 Myr  based on
nuclear turn off time (Meynet et al. 1993) and an age of 125+/-8\,Myr
based on the ``lithium depletion boundary'' (Stauffer el al. 1998). NGC\,2516 appears to be slightly older; Meynet et al. (1993) gives an age of 140$\pm$40~Myr. A mean age of 120$\pm$20\,Myr is considered representative of both clusters for the present purposes.

One way of splitting the data in Table 1 between the PMS and MS models
is to consider the rate of change of radius with time predicted by the
BCAH98 model. PMS stars on Hayashi tracks show a linear decrease in
$\log R$ with $\log t$ (see Fig.~1), whereas the radii of stars once on
the MS is effectively independent of time. For stars aged $\sim$~120\,Myr
the break point between the two regimes occurs at $M
\sim$0.4\,$M_{\odot}$. A more fundamental approach is to consider the
age at which stars produce a significant proportion of their losses by
nuclear burning, hence altering their response to starspots, 
and the time taken for the radius of the star to adapt
to this change, represented approximately by the thermal timescale of the envelope, $t_{\rm env}$ at this point, defined as the  thermal energy contained in the envelop divided by the luminosity of the star (SW86). Table~3 shows estimates of the age at which 50 per cent of the luminosity is provided by hydrogen burning, $t_{\rm nuc}$, and $t_{\rm   env}$ at this age, where $t_{\rm nuc}$ and the depth of the envelope are taken from the models of Siess et al. (2000, for Y=0.277, Z=0.02).  and $t_{\rm env}$ is calculated for the simple case of a polytropic star with the equation of star of a perfect gas (i.e.  n=1.5). Assuming that the starspot coverage remains roughly constant with time then:
\begin{enumerate}
\item Stars with age $t< t_{\rm nuc}$~are expected to exhibit the higher radius inflation characteristic of PMS stars on their Hayashi tracks.

 \item Stars with age $t_{\rm nuc} < t  < t_{\rm nuc} + t_{\rm env}$ are
   expected to show intermediate levels of radius inflation, relaxing on
   a timescale $t_{\rm env}$ from their larger PMS radii towards a
   smaller inflation factor on the MS.

 \item Stars with $t> t_{\rm nuc}+ t_{\rm env}$ are expected to be best
   represented by the radius inflation predicted for MS stars.
\end{enumerate}
  From this, we suggest that stars of age 120$\pm$20 Myr  and
  $M\leq 0.5M_{\odot}$ are expected to show radius inflation characteristic
  of PMS stars, whereas stars of $M > 0.5M_{\odot}$  have effectively
  reached the MS. This split is somewhat imprecise and a little
  model-dependent, however it highlights the importance of $t_{\rm env}$
  in determining the increase in
  radius of spotted stars as they arrive on the MS. Lower mass stars with deep convective zones should retain the radius inflation characteristic of PMS stars long after the inception of nuclear burning.

\begin{table}
\caption{Estimates of the age at which nuclear burning contributes 50
  per cent of the luminosity, $t_{\rm nuc}$ (from the evolutionary model of Siess et al. 2000) and the themal timescale of the envelope
  timescale, $t_{\rm env}$, at this age.}
\centering
\begin{tabular}{ccc}
Mass (M$_{\odot}$) & $t_{\rm nuc}$ (Myr) & $t_{\rm env}$ (Myr) \\ \hline
0.1 & 376 & 894 \\ 
0.2 & 162 & 306 \\
0.3 & 123 & 238 \\
0.4 & 104 & 107 \\
0.5 & 88  & 50 \\ 
0.6 & 75  & 24 \\ \hline
\end{tabular}
\label{tab3}
\end{table}

\begin{table*}[t]
\caption{ The mean increase in radius and estimated filling factor for (i) binary stars,  (ii) higher luminosity stars in the  Pleiades and NGC\,2516 and (iii) lower luminosity stars in NGC\,2516. Filling factors for the first two groups are estimated using a MS model extrapolated from the results of SW86. The filling factor for the third group are estimated using the PMS model described in Section 2.}

\begin{tabular}{lccc}
Sample & Binary Stars with & Pleiades and & NGC2516 \\ 
 & Hipparcos distance & NGC2516 &  \\ \hline
Luminosity & $-0.3>\log L/L_{\odot}>-1.5$ & $-0.3>\log L/L_{\odot}>-1.5$ & $-1.5>\log L/L_{\odot}>-2.3$ \\ 
Model  used & MS & MS & PMS \\ 
Fractional radius increase, $\Delta R/R$ & 0.08 +/- 0.01 & 0.13 +/- 0.03 & 0.40 +/- 0.04 \\ 
Filling factor dark spots, $\beta$ & 0.35 +/- 0.03 & 0.45 +/- 0.07 & 0.51 +/- 0.04 \\ 							
\hline
\end{tabular}
\label{tab4}
\end{table*}

\begin{figure*}
	\centering
  \begin{minipage}[t]{0.95\textwidth}
  \centering
	\includegraphics[width = 130mm]{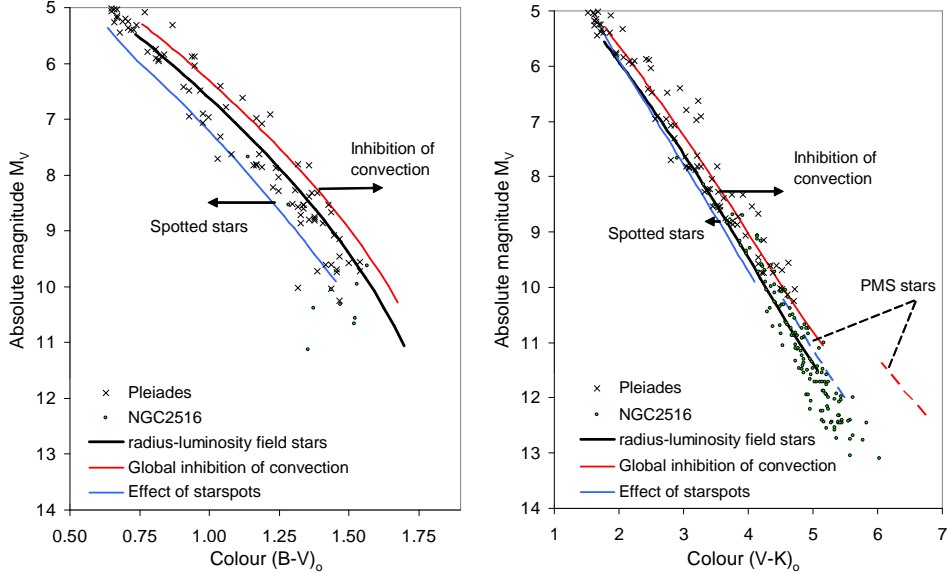}
	\end{minipage}
	\caption{The effect of radius inflation and of starspots on the $M_V$~vs~$(B-V)_0$ and $M_V$~vs~$(V-K)_0$ CMDs. Points show measured data for highly active stars in NGC\,2516 and the Pleiades. The central curves in black shows the empirical radius-luminosity curve for MS field stars corrected to an age of 120\,Myr and [M/H]=0. The upper curves in blue show the effects a increase in radius due to magnetic inhibition of convection, 10 percent for MS stars and 30 percent for PMS stars. The lower curves in red show the effects of 50 percent filling factor of starspots with a spot temperature ratio of 0.7 (giving $\beta=0.4$).} 
\end{figure*}

In Fig.~9 the measured radii in Table 1 are compared with the predicted
radii of stars with $\beta = 0, 0.2, 0.4$ and 0.5 for the MS model
(left) and the PMS model (right). The small open circles on these
models indicate corresponding masses estimated from the BCAH98
model. Based on these values the mean radii for the two higher
luminosity bins of the NGC~2516 sample and all of the Pleiades stars
(see Table~1) have luminosities corresponding to $M> 0.5M_{\odot}$ and
are therefore compared to the MS models in the left hand panel. The
least luminous three bins in the NGC~2516 sample have $M<0.5M_{\odot}$
and are compared to the PMS model. Finally, the radii of the eclipsing
binary components are compared to the MS model, as these are
assumed to be older stars, with
their rapid rotation attributable to tidal locking.
  
Compared to the relationships defined by inactive field stars the
highly active stars in NGC~2516 and the Pleiades show a large radius
inflation. As a percentage, this inflation is largest for low-mass
NGC~2516 stars still in the PMS regime, but is still significant in the
MS regime for stars in NGC~2516, the Pleiades and the MS binary
components. Table 4 shows the weighted mean radius inflation,
$\overline{(R_s-R_u)/R_u}$, for three groups of stars: binary star
components (Table 2), stars of the Pleiades and NGC\,2516 with $-0.3>
\log(L/L_{\odot})> -1.5$ that are in the MS regime and low-mass PMS
stars in NGC~2516 with $-1.5> \log(L/L_{\odot}) > -2.3$ (Table 1). Also
shown in Table 4 are the weighted mean filling factors of {\it dark}
starspots $\overline{\beta}$ that would be required to produce the
measured radius inflation, interpolated from the model loci shown in
Fig.~9. The quoted uncertainties in $\overline{\beta}$ take account of
both the uncertainty in measured radii and the effect of any
uncertainty in the luminosity on the predicted radius of an
unspotted star. This comparison shows that if the observed radius
inflation is produced {\it solely} by (dark) starspots then this would require
filling factors in the range $0.35< \beta <
0.51$. The actual spot coverage fraction would be even larger for spots of
finite temperature. The feasibility of this hypothesis is discussed in
the next section.

\section{Discussion}
Section 2 described a simple polytropic model of a PMS star with dark
starspots covering a fraction $\beta$  of the stellar surface. The
model is applicable to stars on the Hayashi track if:

\begin{enumerate} 
\item The stars are fully convective, without a radiative or a degenerate core and their luminosity results principally from the rate of change of gravitational potential energy as the star contracts.

\item The effect of starspots is to reduce the mean Rosseland opacity averaged over the surface of the star as, $\overline{\kappa} \propto (1-\beta )$, where $\beta$ is the equivalent filling factor of dark starspots that would produce the same reduction in total luminosity as the actual coverage of starspots at the actual starspot temperatures. 

\item The simple relationships governing the radius inflation and
  change in temperature at a given luminosity (Eqn.~10) assume that 
  the starspots were formed at a much earlier
  time. This condition seems likely to be met in clusters like
  NGC~2516 and the Pleiades, because evidence for starspots
  (e.g. modulation of light curves) is routinely reported in much
  younger star forming regions (Herbst \& Mundt 2005; Irwin \&
  Bouvier 2009).

\end{enumerate}
The model is therefore restricted to a range of masses and ages which
must be determined by reference to an evolutionary model for unspotted
stars. In this paper the polytropic model is restricted to stars of
mass $0.2 < M/M_{\odot} < 0.5$ for ages $\leq 120$\,Myr. 

In sections 3 and 4 the predicted increase in radius at a given
luminosity as a function of spot coverage is compared with the measured
radii of active K- and early M-dwarfs in NGC~2516 and the Pleiades and
with the radii of short-period eclipsing binaries with known
distances. In our work we have used the polytropic model to scale an
empirical relationship between radius and luminosity determined from
inactive (and presumably unspotted) field stars with small corrections
for differences in age and metallicity. The use of an empirical
relation is important; at low luminosities the empirical
radius-luminosity relation shows radii 8 per cent higher than the
BCAH98 model isochrones (see Fig.~5). If the radii of active stars were
compared directly with these model radii it would lead to the inference of an
erroneously high level of radius inflation for the lower mass stars.

As an example, Macdonald \& Mullan (2013) rejected starspots as the
sole cause of radius inflation in the same sample of NGC~2516
stars, on the basis that the lowest luminosity objects would require an
excessive ($\beta$ =0.79) level of spot coverage {\it relative to their
  model} of magnetically inactive stars. However, we showed in Fig.~5
that this model significantly underpredicts the interferometrically measured radii of
inactive stars by about 15 per cent. Our lower estimate of the required spot coverage
($\beta = 0.51\pm 0.04$) arises principally because of our
comparison with an empirical radius-luminosity relation.
Using the empirical radius-luminosity relation
does not avoid all systematic uncertainties. 
There is a radius uncertainty of 2--3 per cent in the mean baseline
empirical relationship that can only be improved by more and better
measurements of the interferometric radii of inactive low-mass
stars. However, this level of systematic uncertainty is 
small compared to the radius inflation inferred from the
projected radii of very active low-mass stars (see Fig.~9
and Table~4).

The level of spot coverage our model requires to solely 
account for the inflated radii
of active stars is $\beta=0.35$--0.51. This is a little higher than
the effective filling factors, $\beta$=0.13--0.41, determined
for very active G and K-dwarfs from analysis of their TiO absorption
bands by O'Neal
(2006) but not significantly so. It is still largely unknown what spot
coverage fractions and spot temperatures might exist on very active
M-dwarfs or low-mass PMS stars, though a small value of $\beta$ seems unlikely
given the very high filling factors of kilogauss magnetic field
suggested by Zeeman measurements of M dwarfs with $\log N_{R} < -1$ 
(Reiners, Basri \& Browning 2009). Jackson \& Jeffries (2013)
have shown that the generally small amplitude of $I$-band light curves
of the low-mass NGC~2516 members 
is quite consistent with $\beta \simeq 0.5$ if the spots are small and
scattered over a large portion of the stellar surface. 

The
required spot coverage could be (much) lower if the inflated radii were
also partly (or mostly) explained by the inhibition of convection due
to a globally pervasive magnetic field (Macdonald \& Mullan 2013;
Feiden \& Chaboyer 2012b, 2013).
A relatively simple way to test the relative merits of the starspots
versus globally inhibited convection scenarios is
 to compare the location of active stars and inactive
 field stars in various
colour-magnitude diagrams (CMDs).
Stauffer et al. (2003) pointed out that K-dwarfs in the Pleiades are
nearly 0.5 mag sub-luminous compared to a MS isochrone in the M$_V$  vs
(B-V)$_0$ CMD and also significantly bluer (or sub-luminous) compared to much less active
members of the older Praesepe cluster. However, in the M$_V$ vs (V-K)$_0$
CMD the Pleiades members appear redder (or more luminous). How does
this compare with the anticipated effects of magnetic inhibition of
convection and/or starspots? Expansion driven by magnetic inhibition of
convection would lead to a uniformly lower surface temperature (at a
given luminosity) and should exclusively redden highly active stars relative to magnetically
inactive stars in {\it both} CMDs. 
The effect of starspots depends on the spot temperature ratio. For
darker starspots (with a temperature ratio $\le$ 0.8), the $V$-band flux
density from the spotted area is a relatively small fraction (10 to 20
per cent) of the flux density from the unspotted photosphere. In this
case the effect of starspots will be to shift magnetically active
stars blueward in the $M_V$~vs $(B-V)_0$ CMD. The extent of reddening, if any,
in the  $M_V$~vs $(V-K)_0$  plane is less easy to predict since the
spotted area would still make a significant contribution to the $K$-band flux.

Figure 10 shows CMDs for our sample of fast rotating
stars in NGC~2516 and the Pleiades compared to empirical
colour-magnitude relations for magnetically inactive MS stars from
BM12, corrected to the metallicity and mean age of the clusters. 
Results are transformed from the radius-luminosity relations of BM12 to
the colour-magnitude plane using their $(B-V)_0$ and $(V-K)_0$ colour-temperature relations and the bolometric corrections in Fig.~8. Also shown in Fig.~10 are the predicted effects of:
\begin{itemize}
	\item Magnetic inhibition of convection producing a uniform
          reduction in surface temperature. Curves are shown for a 10
          percent increase in radius for MS stars and 30 per cent increase for PMS stars. 
		\item A simplified, two temperature, model of a spotted
                  star with 50 per cent coverage of starspots, assuming a uniform spot temperature ratio of 0.7 between the spotted and unspotted photosphere, giving $\beta$=0.40. Bolometric corrections for the spotted areas, which are outside the range of the colour-temperature relations of BM12, are taken from the BT-Settl model atmospheres (Allard et al. 2003). 	
\end{itemize}
	
Results in the $M_V$~vs $(B-V)_0$ plane support the interpretation of
Stauffer et al. (2003), that stars in the Pleiades are blue shifted
relative to older, less magnetically active, stars because of
significant spot coverage, though the blueward shift of the Pleiades
stars is less than produced by a spot model with $\beta=0.4$. However a
model solely invoking global inhibition of
convection to explain radius inflation appears to be ruled out 
by the observations.
Results in the $M_V$~vs $(V-K)_0$ are less clear cut. The model of a
spotted star (with spot temperature ratio of 0.7) predicts a small
blueing of active {\it MS stars} relative to inactive MS stars, 
whereas active MS stars in the Pleiades actually show a slight
reddening. However, for the PMS regime at lower luminosities, and with
more radius inflation, the two
models make drastically different predictions. The low-mass PMS stars
in NGC~2516 lie close to the empirical magnetically inactive locus and
this is in reasonabe agreement with the prediction of the $\beta=0.4$
spot model. However to explain a $\geq 30$ per cent radius inflation
seen in these stars using global magnetic inhibition requires much
lower surface temperatures at a given luminosity and results in a
predicted locus that is {\it much} redder than observed. This either
points to (unknown) problems in the indirect methods used for determing
the radii of these low-mass PMS stars or indicates that global
suppression of convection cannot be the sole cause of radius inflation
and favours the starspot model.

It should be stressed that the modelled effects of starspots on the
CMDs are qualitative in nature being based on a simple two temperature
description 
of a spotted star. The real situation is likely to be more complex with
spots and associated plages producing a range of surface temperatures.

\section{Summary}

We have used a simple polytropic model to investigate the effect 
of dark starspots on the evolution and structure of magnetically active 
PMS stars. This model has been combined with an
equivalent treatment of spotted main sequence stars by Spruit \& Weiss
(1986) to predict the level of radius inflation experienced by spotted
stars as a function of their luminosity and spot coverage. The model
has been compared with radii determined for magnetically active 
low-mass PMS stars in NGC~2516, for higher mass stars on or near the ZAMS in the
Pleiades and NGC~2516, and for the tidally locked low-mass components of older
eclipsing binary systems. Our results and conclusions can be summarised as follows:
\begin{enumerate}
   \item Starspots inflate the radii of PMS stars and slow their
     descent along Hayashi tracks. Radii are increased over unspotted
     stars of the same luminosity and age by a factor of $(1-\beta)^{-N}$,
     where $\beta$ is the fractional coverage of dark spots and $N
     \simeq 0.45\pm 0.05$. The temperature of unspotted regions of the
     photosphere is almost unchanged.
     \item For a given $\beta$, the effect of starspots on PMS stellar
       radii is $\sim 2$ times greater than that predicted by Spruit \&
       Weiss (1986) for main sequence stars of similar mass and
       luminosity, where $N \sim 0.2$--0.3 (see Fig.~4).
   \item We find that highly magnetically active K- and M-dwarfs in the
     young Pleiades and NCG~2516 clusters and the tidally-locked binary
     low-mass components show a significant radius
     inflation relative to an empirical locus defined by 
     inactive MS field stars; the mean increase in radius, at a given
     luminosity, ranges from $13
     \pm 3$ per cent for MS K-dwarfs to $40 \pm 4$ per cent for the lower
     luminosity PMS M-dwarfs which are still on their Hayashi tracks.
	\item The observed radius inflation could be caused by magnetic
          inhibition of convection or a high coverage of dark starspots
          or some combination of the two. If starspots are the sole
          cause of radius inflation, this requires a high coverage
          ($0.35 < \beta <0.51$) of dark starspots.
	\item When compared to inactive main sequence stars, the loci
          of the highly active and inflated cluster stars in $(B-V)/V$
          and $(V-K)/V$ colour-magnitude diagrams are more consistent with 
          a high filling factor of dark starspots than an overall
          reduction of temperature caused by the suppression of
          convection. This evidence is strongest for the lowest mass
          PMS stars, where observed $V-K$ colours are far too blue to be
          explained by a globally reduced temperature if radii are inflated by
          30--40 per cent.
\end{enumerate}

\section*{Acknowledgements}

RJJ wishes to thank the UK Science and Technology Facilities Council for
finacial support.

\nocite{LopezMorales2007a}
\nocite{Jackson2009a}
\nocite{Jackson2012a}
\nocite{Mullan2001a}
\nocite{Feiden2012a}
\nocite{Feiden2013b}
\nocite{Spruit1986a}
\nocite{Spruit1982a}
\nocite{Chabrier2007a}
\nocite{Hall1972a}
\nocite{Eaton1979a}
\nocite{Collier1994a}
\nocite{Strassmeier2002a}
\nocite{Marcy1982a}
\nocite{JohnsKrull1996a}
\nocite{Semel1989a}
\nocite{Donati1997a}
\nocite{Stauffer2003a}
\nocite{Stauffer1998a}
\nocite{ONeal1998a}
\nocite{ONeal2004a}
\nocite{ONeal2006a}
\nocite{Boyajian2012a}
\nocite{Prialnik2000a}
\nocite{Hayashi1962a}
\nocite{Boyajian2012b}
\nocite{Siess2000a}
\nocite{Jackson2009a}
\nocite{Jackson2010a}
\nocite{Jackson2010b}
\nocite{Terndrup2002a}
\nocite{Rieke1985a}
\nocite{Hartman2010a}
\nocite{Pinsonneault1998a}
\nocite{An2007a}
\nocite{Allard2003a}
\nocite{Soderblom2009a}
\nocite{Meynet1993a}
\nocite{Leggett1992a}
\nocite{Torres2010a}
\nocite{Rozyczka2013a}
\nocite{Pizzolato2003a}
\nocite{Noyes1984a}
\nocite{MacDonald2013a}
\nocite{Alexander1994a}
\nocite{Baraffe1998a}
\nocite{vanleeuwen2007b}
\nocite{Cutri2003a}
\nocite{Feiden2013a}
\nocite{Jeffries2007a}
\nocite{Krause1980a}
\nocite{Morales2009a}
\nocite{Torres2013a}
\nocite{Jeffries2011a}
\nocite{Irwin2009a}
\nocite{Herbst2005a}
\nocite{Reiners2009a}
\nocite{Dotter2008a}
\nocite{Carpenter2001a}
\nocite{Jackson2013a}
\nocite{Reinhold2013a}

\bibliographystyle{mn2e} 
\bibliography{references}


\bsp 

\label{lastpage}

\end{document}